\documentstyle[11pt,newpasp,twoside]{article}
\input{psfig}
\def\HI{H{\,\small I}}
\newcommand{\ltsima} {$\; \buildrel < \over \sim \;$}   
\newcommand{\gtsima} {$\; \buildrel > \over \sim \;$}
\newcommand{\lta} {\lower.5ex\hbox{\ltsima}}
\newcommand{\gta} {\lower.5ex\hbox{\gtsima}}
\newcommand{\msun}{{$M_\odot$}}
\newcommand{\kms}{$\,$km$\,$s$^{-1}$}

\def\edcomment#1{\iffalse\marginpar{\raggedright\sl#1\/}\else\relax\fi}
\reversemarginpar

\begin{document}
\title{HI on large and small scales in {\sl starburst} radio galaxies}
\author{Raffaella Morganti, Tom Oosterloo}
\affil{Netherlands Foundation for Research in Astronomy, Postbus 2,
NL-7990 AA, Dwingeloo, The Netherlands}
\author{Clive Tadhunter}
\affil{Dep. Physics and Astronomy,
University of Sheffield, S7 3RH, UK}
\author{Bjorn Emonts}
\affil{Kapteyn Astronomical Institute, RuG, Landleven 12, 9747 AD,
Groningen, NL}
\author{Gustaaf van Moorsel}
\affil{National Radio Astronomy Observatory, Socorro,
             NM 87801, USA}

\begin{abstract}
The study of the optical continuum of radio galaxies shows that about
30\% have a young stellar population component. Among them are the
most far-IR bright radio galaxies.  A further indication of the
relatively gas rich environment of these galaxies (possibly related to
the recent merger from which they originate) is the high fraction
being detected in \HI. \\ Here we present recent results obtained from
the study of neutral hydrogen (detected either in emission or
absorption) in a group of starburst radio galaxies.  In some objects,
large-scale (tens of kpc) structures involving \HI\ masses exceeding
10$^9$ \msun\ are observed.  In these cases, the \HI\ can be used to
study the origin and evolution of these systems and the timescales
involved.  In this respect, the parameters obtained from the study of
the stellar populations and from the \HI\ can be complementary. \\ In
other objects, very broad (\gta 1000 km/s), mostly blueshifted \HI\ is
detected in absorption.  This result shows that, despite the extremely
energetic phenomena occurring near an AGN - including the powerful
radio jet - some of the outflowing gas remains, or becomes again,
neutral. This can give new and important insights in the physical
conditions of the gaseous medium around an AGN.  The possible origin
of the extreme kinematics is discussed.

\end{abstract}

\section{Preamble: {\sl Starburst} radio galaxies}

What we will call  {\sl starburst} radio galaxies in this brief review are
radio sources hosted by galaxies that spectroscopically show a young stellar
population component in addition to the old component typical of elliptical
galaxies.  These objects are now known to represent a significant fraction of
samples of radio galaxies (see below). Their relevance, and the motivation for
our study, has to do with the more general and important question of what is the
origin and evolution of radio galaxies.

From imaging studies of powerful radio galaxies (Heckman et al.\
1986), there is clear morphological evidence that their host galaxies
have  undergone interactions and/or mergers. Because of this, it
has been suggested that these phenomena are in fact responsible for
the onset of the radio sources. In this scenario it is not too
surprising that starformation is also triggered by the same merging  event.
The presence of a young stellar population has an other important
implication: it could represent the origin of the UV excess found to
be typical of radio galaxies compared to normal ellipticals (Lilly \&
Longair 1984) - as alternative to the anisotropic scattering model of
the quasar radiation (see e.g. Tadhunter, Fosbury \& di Serego
Alighieri 1988).

Until not so long ago, only a handful of radio galaxies were known,
from optical spectroscopy, to have a young stellar population in their
nuclear regions (see e.g.\ 3C~321 Tadhunter et al.\ 1996, Hydra~A
Melnick et al.\ 1997).  Given the importance of understanding how
common this is, we have recently undergone systematic studies of the
stellar population in radio galaxies (Tadhunter et al.\ 2002, Wills et
al.\ 2002). Because in radio galaxies key absorption features
(sensitive to the presence of young stellar population) are in most
cases filled in by strong emission lines, the study of the stellar
population is better done using the modelling of the entire continuum
SED (see Tadhunter et al.\ 1996 for more details).

\begin{figure*}
\centerline{\psfig{figure=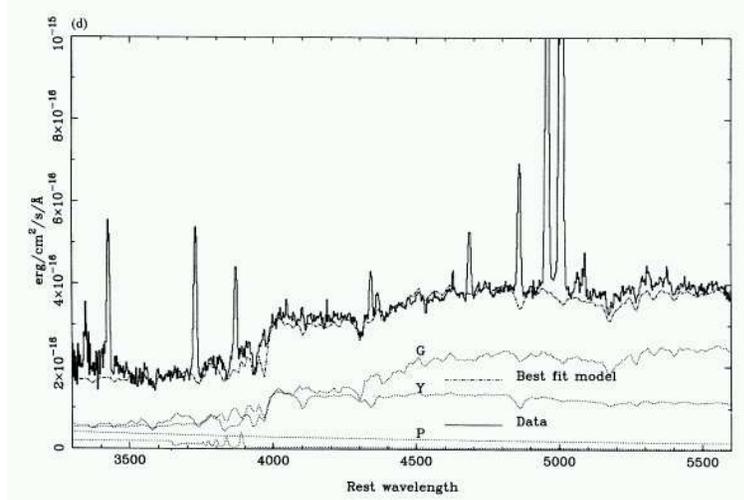,angle=0,width=10cm}
}
\caption{The (rest-frame) intensity spectrum of 3C~321 
(from Tadhunter et la. 1996) with superimposed the model of the continuum -
the best-fitting combination of a 15-Gyr (G), power-law (P), nebular continuum
(N) and a 1-Gyr starburst component (Y). See Tadhunter et al.\ (1996) for
details.}
\end{figure*}

The results show that a young stellar population component is present in at
least 30\% of radio galaxies. The typical ages of the young stars is between
0.5 and 2 Gyr. Interestingly, the galaxies showing this component are also the
most luminous in the far-IR, indicating a link between the optical starburst
and the far-IR emission. This relatively large fraction is also consistent
with the results obtained - using HST - from the UV morphology of powerful 3CR
radio galaxies, where clear regions of starformation are often observed (Allen
et al.\ 2003). {\sl These results support the idea of a merger origin for
radio galaxies and suggest that, at least some of them, had an (ultra-)
luminous far-IR galaxy as progenitor (Tadhunter et al.\ 2002, Wills et al.\
2002)}. The results also indicate that the activity starts late after the
merger event.  The existence of galaxies where this young stellar component is
not observed indicates that, if they are merger related, they are observed at
a later stage and/or they originate from {\sl other type} of mergers. In
summary, the study of {\sl starburst} radio galaxies provides important clues
on the origin and evolution of radio galaxies.

As we will see below, the study of the neutral hydrogen in {\sl
starburst} radio galaxies gives complementary  information
to that provided by the optical stellar population.
Neutral hydrogen has the main advantage that it traces  the large-scale
distribution of the gas as well as the ISM conditions in the nuclear
regions near the AGN. Preliminary results show that in {\sl
starburst} radio galaxies the presence of \HI\ - mainly, but not only,
detected in absorption against the central regions - is more common
compared to other radio galaxies (Morganti et al.\ 2001) and this has
motivated the studies that will be reviewed here.

\begin{figure*}
\vspace{2cm}
\centerline {Figure 2a,b,c in  separate .gif files}
\vspace{2cm}
%\centerline{\psfig{figure=morganti.fig2a.ps,angle=0,width=5cm}
%\psfig{figure=morganti.fig2b.ps,angle=0,width=5cm}
%\psfig{figure=morganti.fig2c.ps,angle=0,width=5cm}
%}
\caption{Examples of \HI\ distribution around ``normal'' early-type 
galaxies obtained from the ATCA follow up of HIPASS detections (see text for details).
Contour levels: $2, 4, 8, 16 \times 10^{19}$ cm$^{-2}$.}
\end{figure*}

\section{Large-scale structures of neutral hydrogen}

\HI\ detected at large radii (tens of kpc) is a  long-lived 
signature of a merger/interaction and therefore provides a key
diagnostic of how a galaxy formed. Large-scale \HI\ structures have
been found in a growing number of nearby ``normal'' (i.e.\ radio
quiet) early-type galaxies (see e.g.\ Oosterloo et al.\ 2002) and in
many cases the likely origin is a major-merger event.  {\sl What is
(if any) the relation between these gas-rich systems and radio
galaxies?} If radio galaxies - and in particular {\sl starburst} radio
galaxies - indeed originate from (major) mergers, and the radio
activity is just a short phase in the evolution of the host galaxy,
fossil large-scale \HI\ structures similar to those detected in
``normal'' early-type might be expected.

A systematic study of the occurrence of large \HI\ structures in radio
galaxies, and {\sl starburst} radio galaxies (to be compared with the
results for normal galaxies of similar type) is now in progress
(Emonts et al.\ in prep).  The results may also shed some light on the
importance of the initial conditions and/or environment for the
evolution of the galaxy.  For example, it has been claimed that the
capability of a merger to bring gas into the nuclear (pc) regions -
and therefore produce an AGN - depends on the structure (e.g.\ the
presence of a bulge) of the progenitor (Mihos \& Hernquist 1994).

Although we do not have yet  complete statistics, here we present  the results
for two interesting cases of {\sl starburst} radio galaxies that we have
recently studied.  First, however, lets summarise briefly what has been found
so far for "normal" early-type galaxies.

\subsection{Large-scale \HI in ``normal'' early-type galaxies: a brief
summary}

 A recent systematic study of \HI\ in all southern RC3 early-type
 galaxies detected by the HIPASS survey (see Oosterloo et al.\ 2003
 for details), has brought to some surprising and interesting results.
 Among the detected galaxies (between 6 and 14 \% depending on the
 optical classification) a wide range of \HI\ morphologies is found,
 but a surprisingly large fraction shows huge structures (many
 tens of kpc and up to 200 kpc) in the form of regularly rotating
 disks containing a large amount of \HI\ ($> 10^9$ \msun).  Some
 examples are shown in Fig.~2. Similar structures previously observed
 in other early-type galaxies (Morganti et al.\ 1997, Oosterloo et
 al.\ 2002) were explained as originating from a major-merger event
 involving at least one gas-rich disk galaxy.  Given that the \HI\
 appears often quite settled, the gas must have already completed at
 least few orbits ($\sim 10^9$ yr) and the merger must be relatively
 old.  Actually, for the huge \HI\ structures shown in Fig.\ 2, a few
 orbits in the outer regions of the disk already imply extremely long
 timescales (several times $10^9$ yr), that may not be always easy to
 reconcile with the merger hypothesis. From the observed parameters a
 ``first-order'' evolution sequence can be build to connect these
 objects (e.g.\ NGC~5266 Morganti et al.\ 1997) with well-known major
 mergers that are observed in an initial phase (e.g.\ NGC~7252 or the
 Antennae; Hibbard \& van Gorkom 1996, Hibbard et al.\ 2001
 respectively).

\subsection{Starburst radio galaxies and large scale \HI}

There are only an handful of very nearby radio galaxies where
large-scale \HI\ structures have been detected so far: for example PKS
B1718-649 (V\'eron-Cetty et al.\ 1995), Centaurus~A (Schiminovich et
al.\ 1994 and refs therein) and, at lower radio power, NGC~4278
(Raimond et al.\ 1981) and NGC~1052 (van Gorkom et al.\ 1986).  In
addition to these, Coma A was the first case where in a powerful radio
galaxy a large \HI\ structure was detected in absorption (Morganti et
al.\ 2002a).  It is interesting that in our study of \HI\ in starburst
radio galaxies we have found already two more interesting cases of
extended
\HI\ - one detected in emission and one in absorption.

\begin{figure*}
\centerline{\psfig{figure=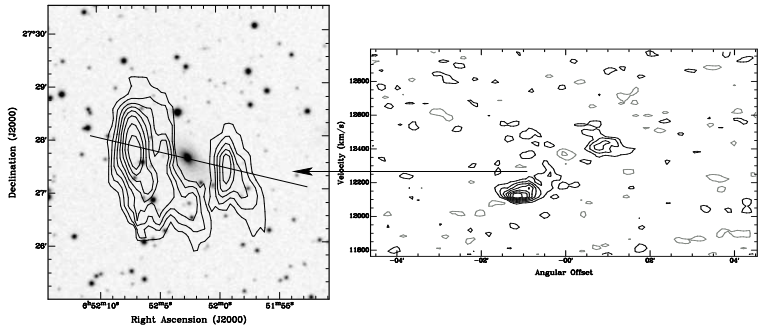,angle=0,width=10cm}}
\caption{{\sl Left:} \HI\ total intensity contours (from the WSRT) of
the  radio galaxy B2~0648+27 superimposed on to an optical image.  
{\sl Right:} Position-velocity plot along the major axis.}
\end{figure*}

In the nearby ($z = 0.041$) {\sl starburst} radio galaxy
B2~0648+27, \HI\ is detected - using the WSRT - both in emission and in
absorption (Morganti et al.\  2003a).  In emission, we detect a large
amount of \HI\ ($M_{\rm HI} = 1.1 \cdot 10^{10}$ $M_\odot$) distributed in a
very extended disk, or ring-like structure, of about 160 kpc in size, as shown
in Fig.\ 3.  We also detect \HI\ absorption against the central radio
continuum component.  The characteristics of the detected \HI, and the
similarities with some of the ``normal'' elliptical mentioned above, are
explained as the result of a major merger.  The structure of the \HI\ and the
asymmetric density distribution, however, suggest that the \HI\ is not yet in a
completely settled configuration.  We can derive a rough indication of the age
of the merger of at least few times $10^8$ yr and it may
represent a merger in an intermediate stage (Morganti et al.\  2003a): between
major mergers in their early phase - like the Antennae - and galaxies with more
settled \HI\ structures like, e.g.,  NGC~5266 (Morganti et al.\  1997).
This is illustrated in Fig. 4.
Interestingly, the radio source shows all the characteristics of being much
younger (compact and steep spectrum, see O'Dea 1998 for a review), $<< 10^7$
yrs old. This supports the idea that the radio source appears late
after the merger but also that, under favourable conditions, the \HI\ gas can
stay around for very long time.  Indeed, it is interesting to note that in
B2~0648+27, as in the other systems mentioned above, the column density of the
\HI\ detected in emission is relatively low (only $\sim 0.8$
$M_{\sun}$pc$^{-2}$) as in the other gas-rich elliptical galaxies (see e.g.\
Oosterloo et al.\ 2002). Therefore, no significant star formation is, at
present, occurring in the regions coincident with the \HI.  The galaxy can,
therefore, remain gas rich for a very long period.

\begin{figure*}
\vspace{2cm}
\centerline {Figure 4 in a separate .gif file}
\vspace{2cm}
%\centerline{\psfig{figure=morganti.fig4.eps,angle=0,width=16cm}}         
\caption{Possible evolutionary sequence (from Morganti et al. 2003a)
linking gas-rich mergers with radio
galaxies and gas-rich ellipticals (described in text). The images of
the Antennas and NGC~7252 have been taken from Hibbard \& van Gorkom
(2001), B2~0648+27 from this paper, NGC~5266 from the data presented
in Morganti et al. (1997).  }
\end{figure*}

A different case appears to be 3C~433. Extended \HI\ in absorption has
been observed against the southern radio lobe of this {\sl starburst}
radio galaxy.  The preliminary
analysis of the data  (Fig.~5) shows that $\sim 5\times 10^8$ \msun\ of
(extended) \HI\ is detected in absorption at about 60 kpc from the
radio core. The gas shows a velocity gradient, but at the moment it is
not clear whether the detected \HI\ is part of an extended gas
disk/tail or whether it corresponds to a region of interaction between
the ISM and the radio lobe.  Given the smaller amount of \HI\ detected
in this object, the origin of this galaxy might be different from the
major-merger suggested for B2~0648+27.  It is interesting to note
that 3C~433 is part of a dumbbell system (Parma et al. 1991) and has a
relatively young stellar population (0.1 Gyr, Wills et al.\ 2002).

At the moment information on large-scale \HI\ is limited to only two
objects, but our aim is to extend this study to many more radio
galaxies and to be able to identify possible trends.

\begin{figure*}
\vspace{2cm}
\centerline {Figure 5 in a separate .gif file}
\vspace{2cm}
%\centerline{\psfig{figure=morganti.fig5.eps,angle=0,width=10cm}}
\caption{The
 total intensity image (grey scale) of the \HI\ absorption in
3C~433, superimposed  on to the  radio continuum (contours). On the right
is shown the position-velocity plot taken along the line marked on the zoom-in
image in the middle.}
\end{figure*}

\section{HI in the nuclear regions of {\sl starburst} radio galaxies}

\HI\ at small radii ($\sim$ kpc, i.e.\  nuclear
scales) has been detected in absorption against the strong radio
continuum in many radio galaxies in a number of studies
(e.g.\  Vermeulen et al.\ 2003, Morganti et al.\ 2001 and refs therein for
previous studies).  This neutral hydrogen has been often associated with
gas distributed in a nuclear disk or torus. However, recent
observations indicate that such an interpretation cannot always be
applied and the situation can be much more complex (Morganti 2002b and refs
therein).

In this respect, our initial results on \HI\ in the central region of
{\sl starburst} radio galaxies are interesting. In particular, the
combined study of  the ionized and the  neutral gas has
revealed to be crucial for the understanding of the physical
conditions of the ISM in the central regions of radio galaxies.  A
clear example is the southern, {\sl starburst} radio galaxy
PKS~1549-79. In this object, two redshift systems of ionized gas  were found
(Tadhunter et al.\ 2001). The \HI\ absorption has the same redshift as
the low ionisation gas, while the high ionisation gas appears
blueshifted respect to it. The \HI\ absorption is associated
with a cocoon of material surrounding the tiny ($\sim 200$ pc in size)
radio source. The material is believed to be left over from the event
that triggered the radio source.  The highly ionized material is
instead associated with a nuclear outflows.

\subsection{Fast \HI\ outflows}

Even more intriguing is the discovery, in at least  two {\sl
starburst} radio galaxies, that the presence of fast outflows is
associated not only with ionized gas but also  {\sl with neutral}
gas. This finding  gives new and important insights on the physical
conditions of the gaseous medium around an AGN.

A good example is the young, {\sl starburst} radio galaxy,
4C~12.50. The fast \HI\ outflow was detected using the broad band
(20~MHz) system now available at the Westerbork Synthesis Radio
Telescope (WSRT). Due to the much stronger radio continuum of this
radio galaxy, detection of gas at very low optical depth is possible.
4C~12.50 is a particularly interesting object as it is a prime
candidate for the link between ultraluminous infrared galaxies (ULIGs,
Sanders \& Mirabel 1996) and radio galaxies (Evans et al.  1999). The
radio source is confined to a region $<0.1$ arcsec ($\sim 240$ pc) and
has all the characteristics of young radio sources ($<< 10^7$ yr).
The ISM of this radio galaxy is extremely rich: it is the most far-IR
bright radio galaxy and has a high molecular gas mass (Evans et al.
1999).  A ``stratified'' fast outflow of the highly ionized gas, that
is quite similar to that detected in PKS 1549$-$79, has been detected
by Holt et al.\ (2003).  In Fig.~6a is presented the \HI\ absorption
detected with the WSRT.  The absorption appears clearly complex and
{\sl extremely broad}.  The full range of velocities covered by the
\HI\ absorption is $\sim 2000$ \kms, {\sl the broadest detected so far
in \HI}.  The peak optical depth of the broad component is only $\tau
\sim 0.002$ and the column density of the full system of shallow \HI\
absorption (assuming a covering factor is 1) is $\sim 1.7 \times
10^{20} T_{\rm spin}/100K$ cm$^{-2}$.

A similar situation (fast outflow of both ionized and neutral gas) is
also seen in the {\sl starburst} radio galaxy 3C~293 that, although it
is not classified as young, is nevertheless believed to have
recently experienced a restarting radio activity (see also Emonts et
al.\ these Proceedings).  Again using the new broad band (20~MHz)
system available at the WSRT, broad, mainly blueshifted, \HI\
absorption has been detected (see Morganti et al.\ 2003b and Emonts et
al.\ these proceedings for details).  The absorption profile is shown
in Fig.~6b. The broad \HI\ absorption has a full-width at zero
intensity (FWZI) of $\sim 1400$ \kms. It is very shallow, with a
typical optical depth of only $\sim 0.0015$, and, assuming a covering
factor is 1, a column density of the \HI\ is $\sim 2 \times 10^{20}\
T_{\rm spin}/100$ K cm$^{-2}$.  As in the case of 4C~12.50, this gives
a lower limit to the true column density as the $T_{\rm spin}$
associated with such a fast outflow can be as large as a few 1000~K
(instead of 100~K which is more typical of the cold, quiescent \HI\ in
galaxy disks).

These two galaxies are the first examples of radio galaxies where a
fast \HI\ outflow is observed, but more candidates exist among {\sl
starburst} radio galaxies. This result shows that, despite the
extremely energetic phenomena occurring near an AGN - including the
powerful radio jet - some of the outflowing gas remains, or becomes
again, neutral.

\begin{figure*}
\centerline{\psfig{figure=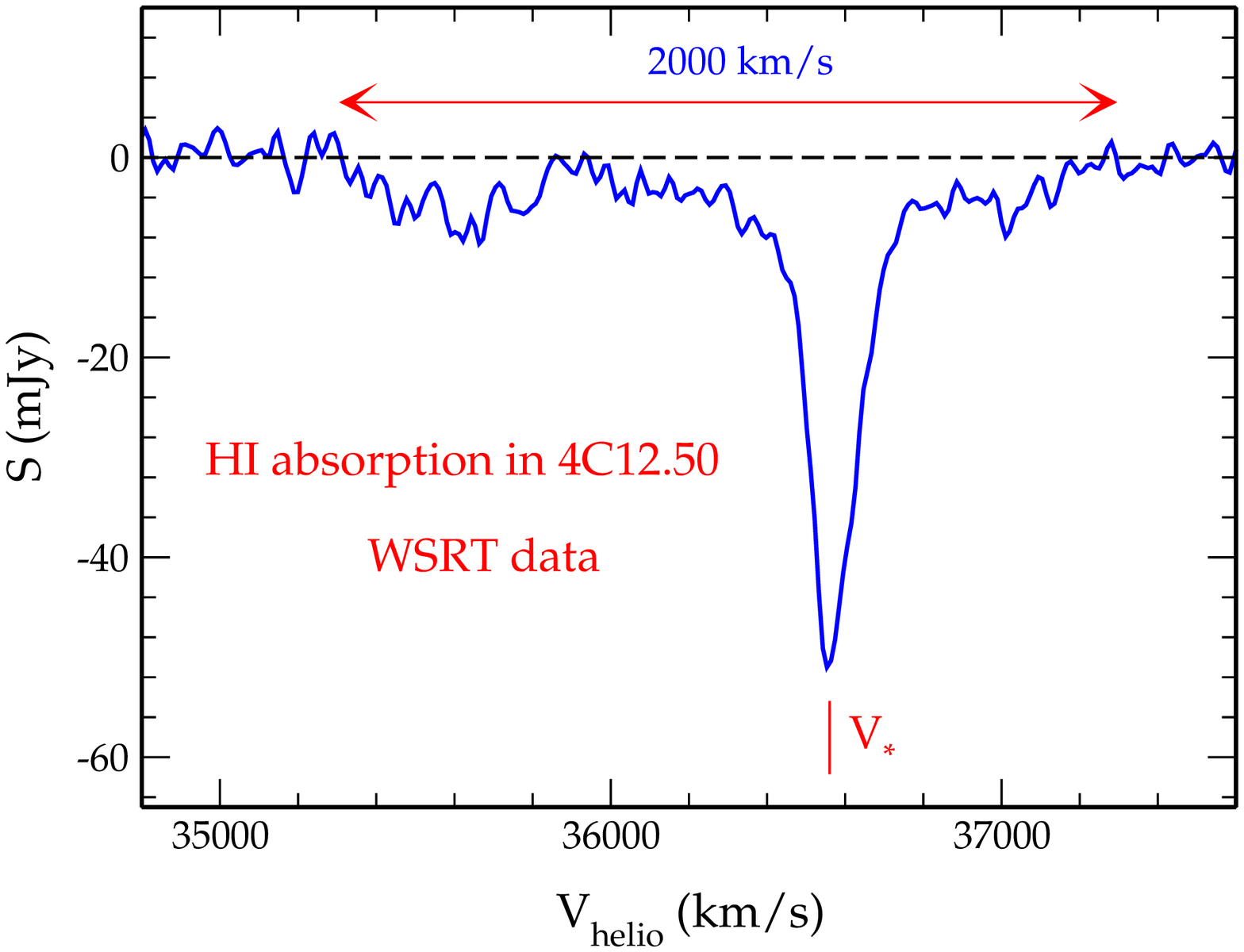,angle=0,width=6cm}
\psfig{figure=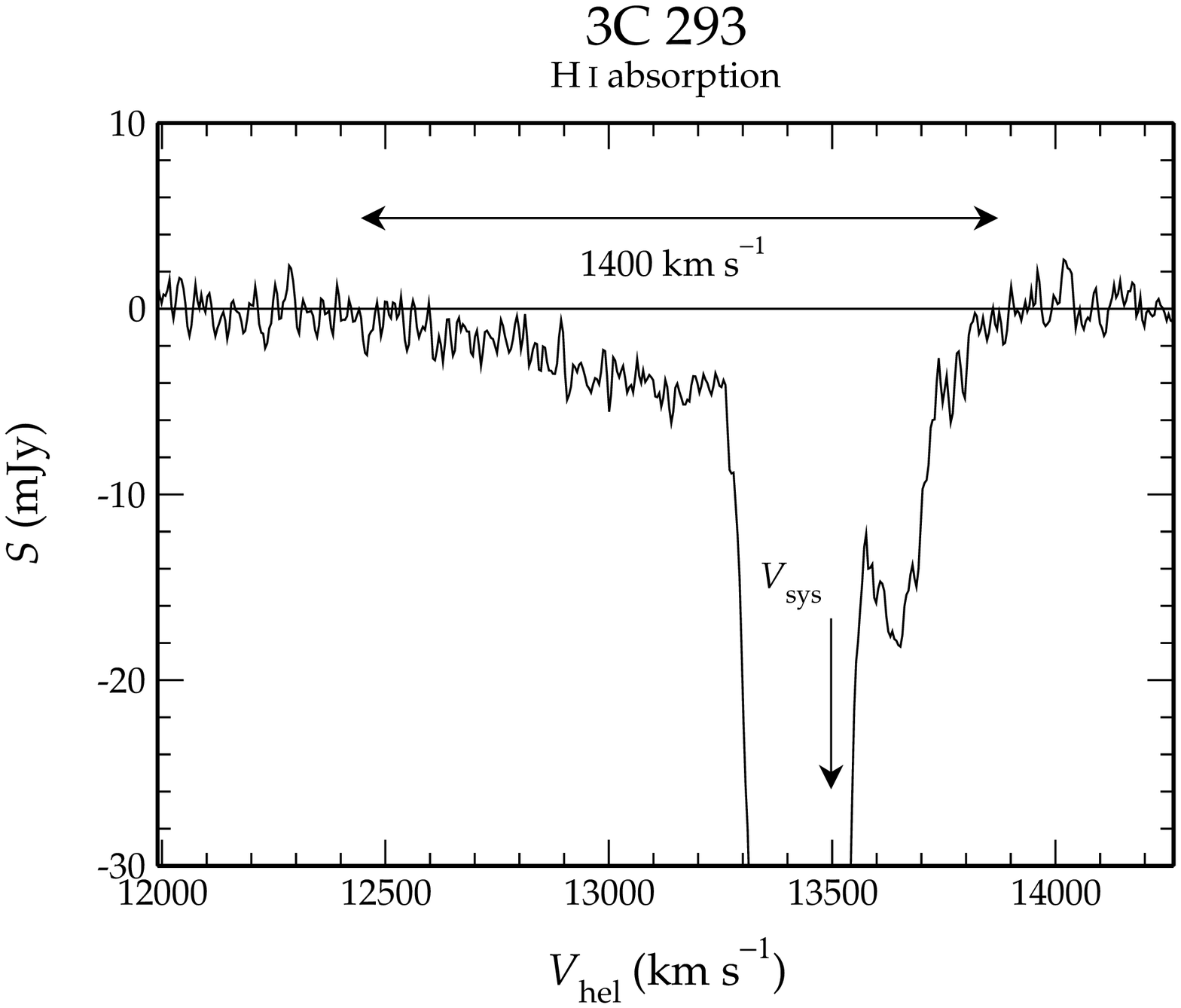,angle=0,width=6cm}} \caption{The \HI\
absorption profile detected in 4C~12.50 (left) and 3C~293 (right) from
the
 WSRT observations.  The spectra are plotted in flux (mJy) against
optical heliocentric velocity in \kms.}\end{figure*}

\subsection{What produces the \HI\ outflows?}

Outflows of ionized gas appear to be a relatively common
characteristic in AGN and starburst galaxies (as also illustrated by a
number of talks in this conference).  Fast gas outflows are now
detected (in optical, UV and X-ray observations) in a wide range of
AGNs, from Seyfert galaxies to quasars (see e.g.\ Crenshaw et al.\
2001; Turnshek 1986; Krongold et al.\ 2003 and refs therein).  These
outflows can be produced by different and highly energetic phenomena,
such as interaction of the radio plasma with the ISM as well as
nuclear and/or starburst winds. It is interesting to see that \HI\
outflows have been also detected in \HI, but so far are only found in
{\sl starburst} radio galaxies.  This might be due to the presence of
a particularly rich ISM that characterises radio galaxies in this
stage of their evolution, with the rich ISM possibly resulting from a
recent merger.

The central question is how {\sl neutral} gas can be associated with
such fast outflows.  A possible model is that the radio plasma jet hits
a (molecular) cloud in the ISM.  As a consequence of this interaction,
the kinematics of the gas is disturbed by the shock and perhaps part
of the gas is also ionized by it.  Once the shock has passed, part of
the gas may have the chance to recombine and become neutral, while it
is moving at high velocities (see Fig.~7). In the model proposed by
Mellema et al.\ (2002), as the shock runs over a cloud, a compression
phase starts because the cloud gets embedded in an overpressured
cocoon. The shock waves start travelling {\sl into} the cloud and the
cloud fragments with the fragments moving at high velocities. The
cooling times for the dense fragments are very short (few times $10^2$
years) compared to the lifetime of the radio source and that the
excess of energy is quickly radiated away. This results in the {\sl
formation of dense, cool and fragmented structures at high
velocities}.

As indirect support to this hypothesis, it is worth mentioning that in
the only other case of broad blueshifted \HI\ absorption (of 700 \kms\
FWZI) studied in detail so far, the Seyfert galaxy IC~5063 (Oosterloo
et al.\ 2000), the \HI\ absorption is coincident with the brighter
radio lobe where also the most kinematically disturbed ionized gas is
observed. This supports the idea of jet/cloud interaction as most
likely mechanism in this Seyfert galaxy.  Another possible example is
the Seyfert galaxy Mrk~1 (Omar et al.\ 2002). However, it is not yet
clear whether with the mechanism proposed above it is really possible
to accelerate the clouds to the high velocities observed.  Therefore,
other possibilities should be considered:

{\sl i) Starburst:} Given the presence of a young stellar population
in these galaxies, a starburst wind capable to accelerate the gas
should be considered to explain the \HI\ outflow. This mechanism is
indeed known to produce fast outflow at least of ionized gas (see
e.g.\ Heckman, Armus \& Miley 1990, and Veilleux these proceedings).
However, in the case of {\sl starburst} radio galaxies the age of the
young stellar population component has been estimated between 0.4 and
2.5 Gyr.  Thus, the neutral outflow would have to be a ``fossil''
starburst-driven wind from the strong starburst that may have occurred
of the order of 1 Gyr ago.  Given this condition, it is not clear
whether the starburst-driven outflow would survive to the present day
and, more important, whether it would be still seen against the
central regions.

{\sl ii) Radiation pressure from the AGN:} Dopita et al. (2002) have
explored the possibility that the narrow-line regions in active
galaxies are actually dusty and that the radiation pressure acting
over the dust could produce the extreme kinematics observed in the
ionized gas.  In their model they consider a dusty (radiation
pressure-dominated) region surrounding a (photoevaporating) molecular
cloud. In this scenario, the dust is estimated to survive as the
average gas temperature does not exceed $10^4$ -- unlike the case of
fast shocks -- and it will not destroy the grains in the short time
during the passage through the high-emissivity region.  While for
4C~12.50 the strong AGN suggest that this mechanism could be 
considered as an option, in 3C~293, the very low ionisation and faint
emission lines suggest that the UV radiation from the AGN is not that
strong and therefore unlikely to provide the necessary pressure to
produce the outflow.

{\sl iii)} A further possibility is that the \HI\ outflow is associated to the
adiabatically expanded broad emission line clouds (BELCs) as suggested by
Elvis et al. (2002).  Following their work, if the broad emission line region
is outflowing, then the BELCs will expand and cool adiabatically, and will
reach 1000~K at $\sim$ few pc. They will then form dust and also neutral
hydrogen but in this case the \HI\ outflow has to be located very close
to the nucleus.

In summary, in order to identify which mechanism is the more likely to
accelerate the gas, the exact location of the broad \HI\ absorption is
needed and VLBI follow-up observations are now in progress.

\begin{figure*}
\centerline{\psfig{figure=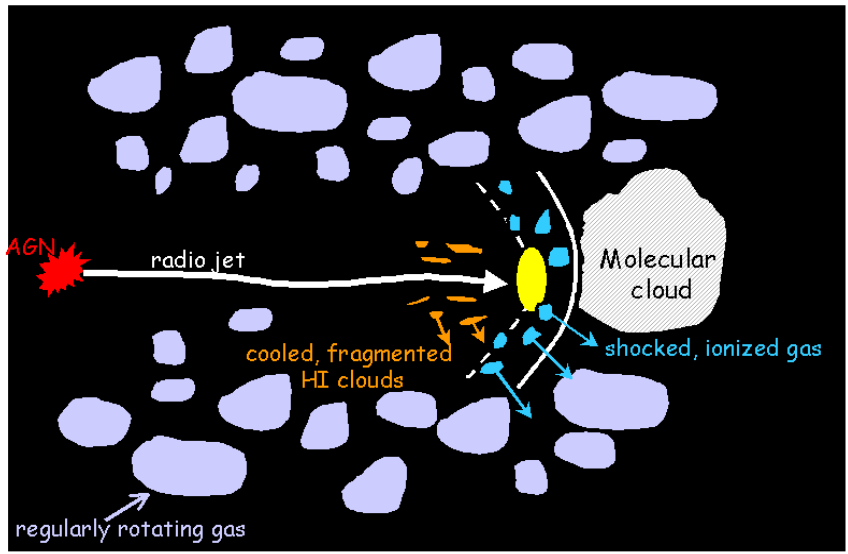,angle=0,width=12cm}}
\caption{A schematic diagram of the model proposed for the {\sl starburst} 
radio galaxies discussed in the test where outflows of neutral and ionized gas
have been detected.}
\end{figure*}

\section{Final remarks}

In this review we have shown that the group of {\sl starburst} radio
galaxies is a particularly interesting one in order to understand the
origin and evolution of radio galaxies. The study of neutral hydrogen
in this group of objects is also giving us important clues about these
open questions. \HI\ appears to be relatively common in starburst
radio galaxies (at least compared to other radio galaxies).
We interpret this as an indication that large amount of gas is still
around from the merger that triggers the radio source.

We have presented two cases in which extended \HI\ has been detected.
The case of B2~0648+27 represents one of the largest \HI\ disks known
($\sim 160$ kpc) around a radio galaxy with more than $10^{10}$\msun\
of \HI.  It shows that at least some radio galaxies are likely the
results of major mergers and that the AGN activity appears late after
the merger.

Even more surprising is the finding of fast
\HI\ outflows in at least two {\sl starburst} radio galaxies. 
This result shows that, despite the extremely energetic phenomena
occurring near an AGN - including the powerful radio jet - some of the
outflowing gas remains, or becomes again, neutral. This can give new
and important insights on the physical conditions of the gaseous
medium around an AGN.  

Following these results, a tantalising connection can be made with the
high-$z$ radio galaxies. As in the {\sl starburst} radio galaxies presented
here, there is clear evidence for the presence of large amounts of cold gas
and, in general, for the presence of a rich gaseous environment in radio
galaxies at high redshift (see e.g.\ van Breugel 2000 for a
review). Particularly relevant is the finding (Villar-Marti\'n et al.\ 2002)
of a low surface brightness Ly$\alpha$ halo with quiescent kinematics in the
case of the high-$z$ radio galaxy USS~0828+193.  One possible way suggested to
explain this structure is that the low surface brightness Ly$\alpha$ halo is
the progenitor of the \HI\ discs as found in the case of B2~0648+27 and others
discussed above. If this is the case, the wealth of details that we can learn
at low-$z$ may be crucial for understanding the structures at high-$z$.
Moreover, strong interactions between the radio plasma and the medium are
expected to be very important in high-$z$ objects.  Outflow phenomena have
been detected in many high-redshift radio galaxies. In many cases, asymmetric
Ly$\alpha$ profiles suggest the presence of blueshifted absorbing gas (likely
neutral hydrogen; see van Ojik et al.\ 1997, de Breuck et al.\
1999). Additionally, complex gas kinematics is also observed in a large
fraction of high-$z$ radio galaxies (van Ojik et al.\ 1997).  Thus, similar
processes as observed in 4C~12.50 and 3C~293 are likely to be even more common
in these high-$z$ systems. Understanding the physics of fast gas outflows and
the conditions for which part of the outflowing gas is neutral, can be also
relevant for understanding high-$z$ objects.

\end{document}